\documentclass{article}
\usepackage{cite}
\usepackage{amsmath,amssymb,amsfonts}
\usepackage{algorithmic}
\usepackage{graphicx}
\usepackage{textcomp}
\usepackage{multirow}
\usepackage[table,xcdraw]{xcolor}
\usepackage{tabularx}
\usepackage{booktabs}
\usepackage{float}
\usepackage{hyperref}
\usepackage[all]{hypcap} 
\usepackage{authblk}
\title{Volumetric Image Projection Super-Resolution Ultrasound (VIP-SR) with a 1D Unfocused Linear Array}

\author{Bingxue Wang, Kai Riemer, Matthieu Toulemonde,Jipeng Yan, Xiaowei Zhou, Meng-Xing Tang
\date{}
\thanks{Correspondence Author: Meng-Xing Tang (mengxing.tang@ic.ac.uk)}
\thanks{B.W., K.R., M.T. J.Y. and M.T are with Ultrasound Lab for Imaging and Sensing, Department of Bioengineering, Imperial College London, London, UK, SW7 2AZ. (e-mail: bingxue.wang18@imperial.ac.uk; mengxing.tang@imperial.ac.uk).}
\thanks{Z.X. is with State Key Laboratory of Ultrasonic Medical Engineering, Chongqing Medical University, Chongqing, China. (e-mail: zhou.xiaowei@cqmu.edu.cn)}}

\begin{document}
\maketitle

\begin{abstract}
Super-Resolution Ultrasound (SRUS) through localizing spatially isolated microbubbles has been demonstrated to overcome the wave diffraction limit and reveal the microvascular structure and flow information at the microscopic scale. However, 3D SRUS imaging remains a challenge due to the fabrication and computational complexity of 2D matrix array probes and connections. Inspired by X-ray radiography which can present volumetric information in a single projection image with much simpler hardware than X-ray CT, this study investigates the feasibility of volumetric image projection super-resolution (VIP-SR) ultrasound using a 1D unfocused linear array. Both simulation and experiments were conducted on 3D microvessel phantoms using a 1D linear array with or without an elevational focus, and a 2D matrix array as the reference. Results show that, VIP-SR, using an unfocused 1D array probe can capture significantly more volumetric information than the conventional 1D elevational focused probe. Compared with the 2D projection image of the full 3D SRUS results using the 2D array probe with the same aperture size, VIP-SR has similar volumetric coverage using 32 folds less independent elements. The impact of bubble concentration and vascular density on the VIP-SR US was also investigated. This study demonstrates the ability of high-resolution volumetric imaging of microvascular structures at significantly reduced costs with VIP-SR.
\end{abstract}


\section{INTRODUCTION}
\label{sec:introduction}
Visualization of microvasculature can assist in the diagnosis, treatment and monitoring of diseases such as cancer with abnormal proliferation of blood vessels or cardiovascular diseases. Studies have demonstrated Super-Resolution ultrasound (SRUS) imaging via separation of two targets with sub-diffraction resolution \textit{in vitro} \cite{Viessmann2013,Desailly2013} and \textit{in vivo} \cite{Christensen-Jeffries2015,Errico2015} by localizing the spatially isolated microbubble or vaporized nanodroplet \cite{Zhang2018,Zhang2019a} signals in successive frames of images. However, most studies are 2D imaging with limited volume coverage using a traditional 1D array probe, while volumetric visualization is valuable as biological tissues are often complex 3D structures.

Previously, volumetric SRUS was done by scanning a 1D probe along the elevational direction to acquire multiple 2D slices of images \cite{Errico2015,Lin2017,Zhu2019}. The mechanical scanning process is slow and super-resolution is only achieved in two of the three spatial directions. SR in all three spatial directions can be achieved using several rows of parallel elements for imaging and then localizing the elevation positions by fitting the parabolas \cite{Desailly2013} or using two orthogonal clinical transducers and imaging at the overlapping region at the focus \cite{Christensen-Jeffries2017}. Customized transducers e.g. a hemispherical transcranial array can be used to visualize the spiral structure of an ex vivo skull \cite{Reilly2013} and the whole brain \cite{Demeulenaere2022}, but the fabrication cost is expensive. The application of a full 2D matrix array probe has also been investigated \cite{Foroozan2018}, but the hardware is complex and expensive as it requires excessive elements. Multiplexed 2D matrix array \cite{Heiles2019} was used to mitigate the problem by reducing the number of acquisition channels from 1024 to 512, at a cost of temporal resolution. Furthermore, to overcome the problem of high channel count and large data volume, a sparse 2D array with elements following a density-tapered 2D spiral layout \cite{Ramalli2015} has been used to generate 3D SR images \cite{Harput2020b}, with slightly degraded initial image quality. Researchers also tried to optimize the transmit sequences to reduce the number of transmit and receive combinations at little cost to SNR \cite{Chavignon2021}. Shared-Sparse apertures \cite{Wang2021} were also demonstrated for improving acquisition efficiency. But a large amount of data still poses challenges in the clinical translation of the 3D SRUS imaging technique. Alternatively, a row-column-addressed array was also shown to be able to generate 3D SR images with far fewer acquisition channels and at the same time larger field-of-view (FOV)  \cite{Jensen2020}, and the coherence beamforming technique has been investigated to reduce the level of edge waves that will help improve the performance of this probe in 3D SR \cite{Hansen-Shearer2022}, but this probe hasn’t been clinically approved.

\begin{figure*}[!b]
\centerline{\includegraphics[width=\textwidth]{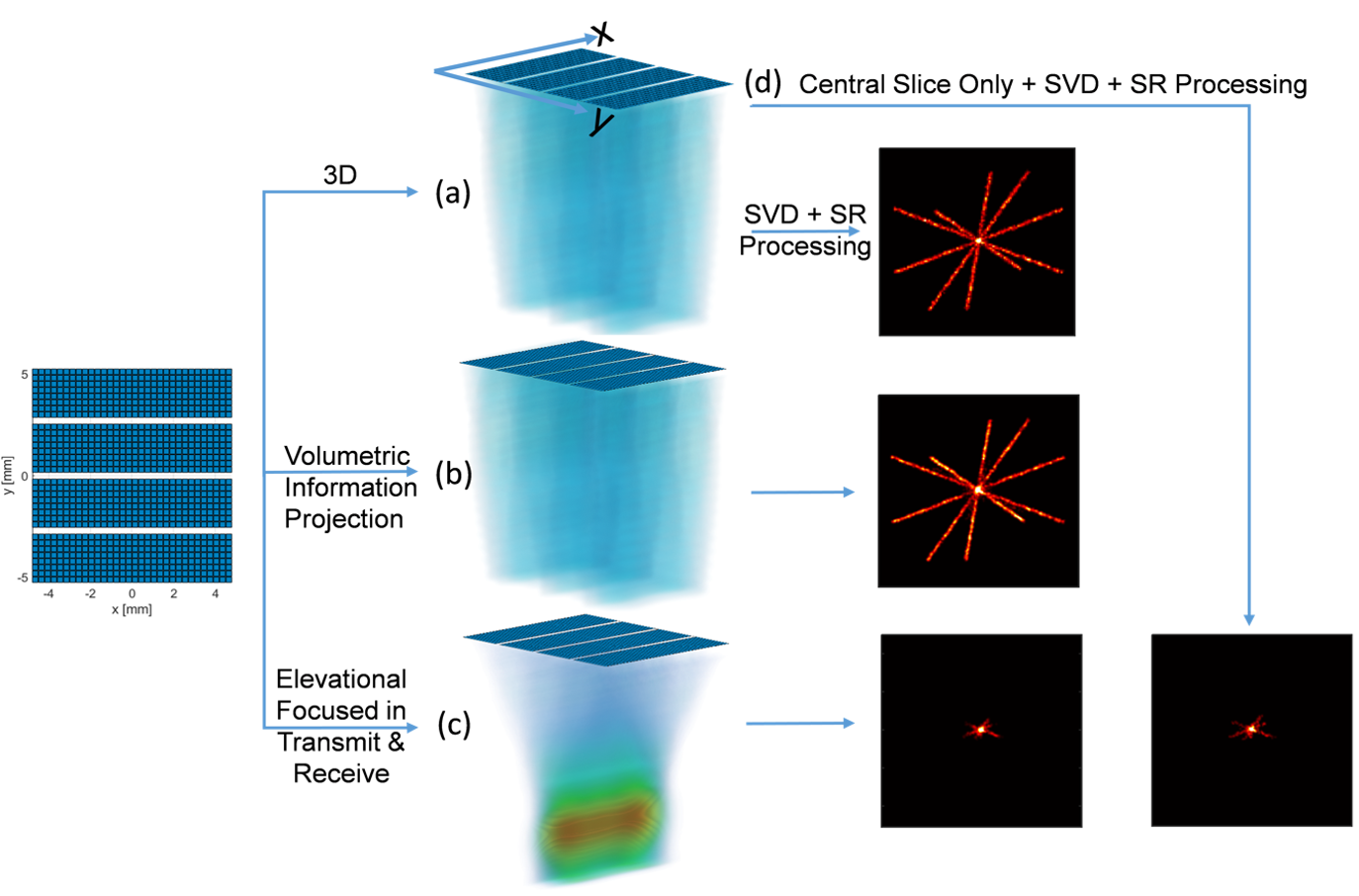}}
\caption{Transducer and four imaging schemes. The diagram of the 32*32 matrix array transducer is shown on the left. (a) 3D imaging through plane wave transmission and echo reception by all 32x32 transducer elements; (b) VIP imaging throughplane wave transmission and echoes received and summed along the elevational direction; (c) Elevational focused (EF) imaging by delaying the transmit and received echoes and sum along the elevational direction of the 32x32 transducer elements; (d) Central slice of the 3D beam formed volume (i.e. similar to (c) but only receive focused)}
\label{fig1}
\end{figure*}

In the UK’s clinical radiology, the most commonly used diagnostic imaging modality is X-ray radiography \cite{NHSEnglandandNHSImprovment2020}. Compared with full 3D X-Ray Computerized Tomography (CT), X-ray radiography has much simpler hardware and lower cost, while still having volumetric information through condensing a 3D volume into a 2D projection image. This study takes inspiration from X-Ray radiography to investigate a potential alternative solution for volumetric information generation and display in SRUS. The aim of this study is to demonstrate the feasibility of volumetric information projection SR (VIP-SR) imaging using a clinically easily available 1D unfocused array. First, a proof-of-concept is shown with a microvessel phantom in simulation, followed by experimental results acquired by a 1D unfocused linear array approached by a 2D matrix probe. The results were compared with the traditional 1D focused linear array and 2D array results. The effect of the microvasculature complexity and microbubble concentrations on its performance was also studied. 

\section{METHOD}

\subsection{Transducers}

To evaluate the proposed VIP-SR, several probes of the same aperture size were investigated, including 1) a 2D matrix array probe; 2) an elevational unfocused 1D array; 3) an elevational focused 1D array. To investigate the three probes, a 2D matrix array aperture was used at first and then data from the other probes were synthesized using data from this 2D matrix array.

Firstly, a full 2D probe was used based on the multiplexing matrix array probe (Vermon, Tours, France) with 1024 elements, a central frequency of 7.8MHz and an aperture size of 9.3 mm x 10.2 mm, with four sub-apertures separated by three empty row lines as shown in Fig. \ref{fig1}. Plane waves were transmitted using all the 32*32 elements simultaneously and the received radiofrequency (RF) data was beam formed using the 3D Delay and Sum (DAS) algorithm. Secondly, to simulate a 1D unfocused linear array with the same aperture size, the RF data acquired by the 2D array probe were summed along one lateral direction, effectively reducing the channel number from 32*32 to 32*1. Thirdly, the simulation of a 1D linear array with an elevational focus was implemented in two ways. One way was to apply transmit delays to the matrix array elements to form an elevational focus and receive time delays before they were summed in the elevational direction, which would simulate a classic 1D linear array with an elevational focus through an acoustic lens. The other was to take the central slice image out of the 3D volumetric B-Mode image, as the beam width at the focal region has the same order of wavelength and constrains the imaging volume. These four schemes were referred to as 3D, VIP (volumetric information projection), EF (elevational focused in transmit and receive) and CS (central slice in this study (see Fig. \ref{fig1}). The x-direction is referred to as the lateral direction, while the y-direction along which the gaps exist is referred to as the elevational direction.

\subsection{Simulation}
The simulation was performed in Field II \cite{Jensen1996} with probe parameters shown in Table \ref{tab1}. The flat aperture can be seen in Fig. \ref{fig1}, where the three blank rows between each sub-aperture indicate the dead rows for the fabrication purpose. Apodization was applied along the lateral direction to keep the elevational profile unaffected.

Microbubbles are simulated as linear point scatters. Two types of point scatter phantoms were used in simulations: single scatters and micro-sized cross-tube phantoms. In the single scatter simulations, only one linear scatter was present at each frame and was initially placed at (x=0, z=20) mm and moved along the elevational direction with a one-wavelength step size (y=-5:0.2:5 mm). PSFs of scatters at different places and the localization errors along both lateral and axial directions were compared between these four different methods. 

\begin{figure}[!b]
\centerline{\includegraphics[width=0.7\columnwidth]{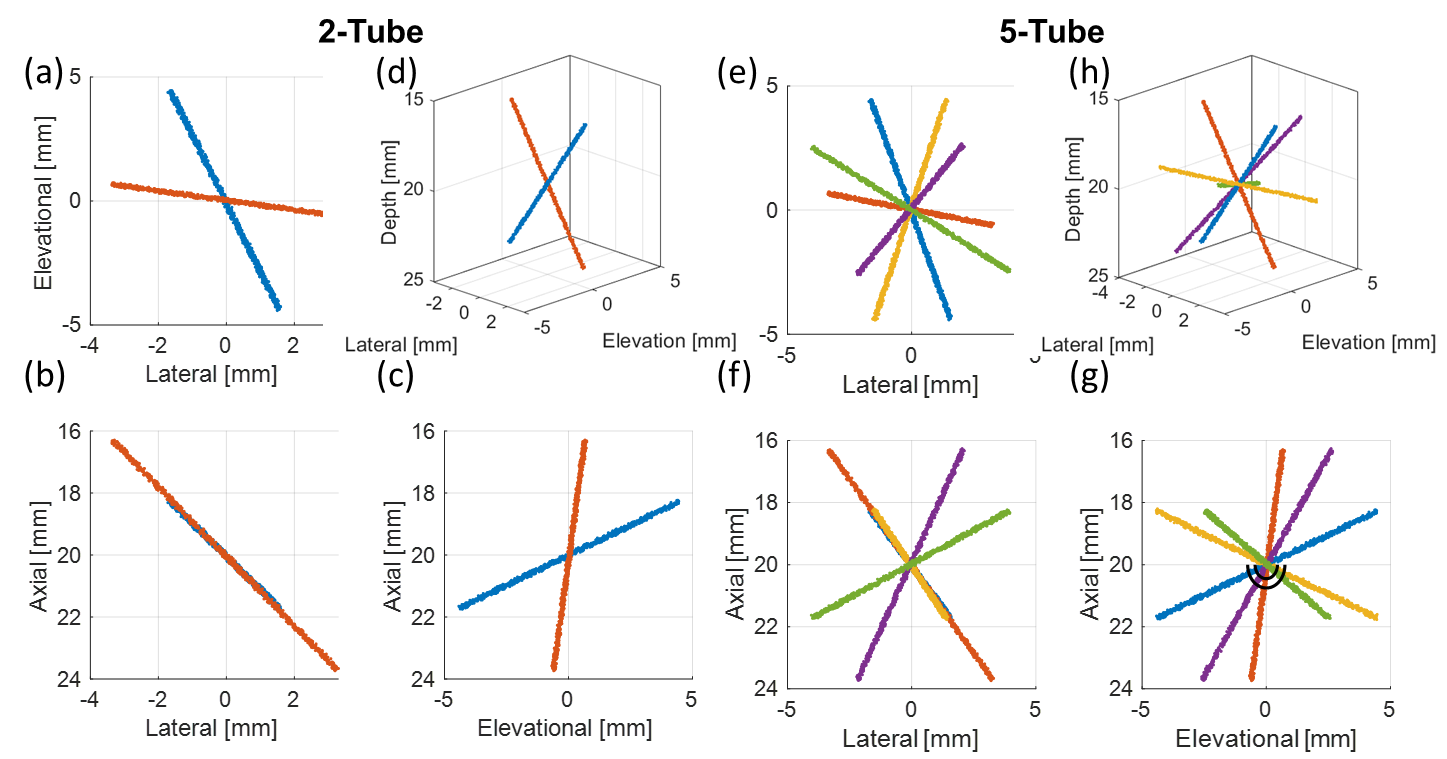}}
\caption{Ground truth of the cross-tube simulation phantoms. (a)-(c) and (e)-(g) show the projection views from each of the three directions and (d) and (h) show the 3D view of the 2-tube and 5-tube phantom, respectively. The black rings indicate the distances which are 350 and 750 $\mu$m away from the origin.}
\label{fig2}
\end{figure}

For the micro-sized cross-tube phantom, a 2-tube structure and a 5-tube structure (Fig. \ref{fig2}) were used to investigate the effect of the vascular complexity, with each of the tubes filled with 1000 scatters randomly over time. The effect of scatter’s concentration was also investigated by tuning the number of scatters per tube per frame from 1 to 5, reducing the total number of frames from 1000 to 200 with the total number of scatters fixed. For the cross-tube phantom simulation dataset, white noise was added to achieve a 3 dB SNR in the RF channel data. For the evaluation, the radial profiles within a half ring-shaped ROI that is labeled in black lines in Fig. \ref{fig2}(g) are measured. The radial profile along different angles within the half-circle was calculated by averaging the single-pixel circular intensity with the neighboring pixels along the radial direction. The localization performance was also evaluated quantitatively by calculating the localization precision, sensitivity and Jaccard Index for different methods, with a localization error tolerance level equals to half of the wavelength. 

\begin{table}[h!]
\caption{Simulation Parameters}
\label{table}
\renewcommand{\arraystretch}{1.3}
\centering
\begin{tabular}{p{100pt}p{100pt}}
\toprule
\textbf{Parameter name} & \textbf{Value}     \\ \midrule
Centre frequency    & 7.8 MHz         \\
Speed of Sound     & 1540 m/s        \\
Wavelength       & 197 $\mu$m         \\
Element width      & 0.275 mm        \\
Element kerf      & 0.025 mm        \\
Element number     & 32*32 (1024)      \\
Aperture size      & 9.6 (x) *10.2 (y) mm  \\
Sampling frequency   & 31.24 MHz        \\
Pulse          & 2-cycles, Hann-weighted \\
Apodization       & Tukey(0.5)       \\
Elevational Focus Depth & 20 mm          \\ \bottomrule
\end{tabular}
\label{tab1}
\end{table}

\subsection{\textit{In vitro} Experiments}

\begin{figure}[!b]
\centerline{\includegraphics[width=0.8\columnwidth]{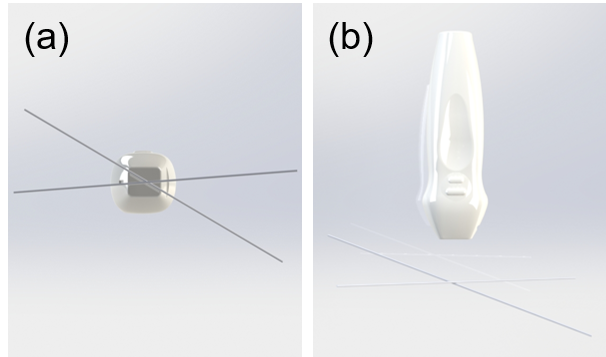}}
\caption{Bottom (a) and side (b) view of the experiment set-up. These two 200 micron microtubes were positioned using a grid of positioning holes on the sidewalls, with acoustic absorbing pad at the bottom of the phantom.}
\label{fig3}
\end{figure}

The 7.8MHz multiplexed matrix probe was used in this study and was driven by a Vantage 256 system (Verasonics Inc., Kirkland, WA, USA). This probe has no physical elevational focus, providing the flexibility to simulate different types of 1D transducer arrays. To reduce the effect of motion between compounding transmissions and switching between different sub-apertures, the transmit sequence was kept as simple as possible. Four sub-apertures were driven at the same time to send plane waves and then each sub-aperture was used to receive the RF signal sequentially, which means that each frame of acquisition consists of four separate acquisitions. The mechanical index (MI) measured in the middle of the probe was 0.05. A total number of 1000 Frames of images were acquired with a frame rate of 500 Hz without any compounding. Two 200 $\mu$m ± 15 $\mu$m Hemophan cellulose tubes (Membrana, 3M, Wuppertal, Germany) with a wall thickness of 8 ± 1 $\mu$m were crossed at a depth of 16 mm, where the edge waves caused by the dead rows between sub-apertures diminish below -14 dB \cite{Harput2020b}. A probe holder was 3D-printed to help align the probe with the phantom. One tube was kept in a central slice of the 3D imaging volume as much as possible, while the other one was tilted and spanned across the 3D space, as shown in Fig. \ref{fig3}. Microbubbles(MBs) were homemade according to the method stated in \cite{Lin2017a}, and the MB solution was diluted by 5000 times, resulting in a concentration of $10^6$ per ml. Then the diluted suspension was injected into the tubes with a constant flow rate of 30 ul/min using a syringe pump (Harvard Apparatus, Holliston, MA, USA), which equals to an average flow speed of 60 mm/s. Volumes were composed of 51 × 101 × 103 voxels with 100 $\mu$m in all three spatial directions.

\subsection{\textit{In vivo} Experiments}
The experiment was authorised by the Animal Welfare and Ethical Review Body of Imperial College London and followed the Animals (Scientific Procedures) Act 1986. A specific-pathogen-free New Zealand White rabbit (male, HSDIF strain, age 13 weeks, weight 2.4 kg, Envigo, UK) was sedated with acepromazine (0.5 mg/kg, i.m.) and anaesthetized with medetomidine (Domitor, 0.25 mL/kg, i.m.) and ketamine (Narketan, 0.15 mL/kg, i.m.). A bolus of 0.1 ml Homemade MBs was injected into the rabbit. Imaging was performed on its left kidney after it was shaved. The same imaging sequence as the \textit{in vitro} experiment was used to collect 1500 frames of images with a frame rate of 500 Hz.

\subsection{Post-Processing}
For the experimental dataset, RF data from four sub-apertures were combined and beam formed to reconstruct a single 3D volumetric image using the 3D DAS algorithm. Then SVD filtering \cite{Deffieux2015} was applied to the beam formed data to remove the strong reflections from the tube walls and tissue signals, where the SVD threshold was chosen based on the spatial correlation method \cite{Baranger2018}. The VIP dataset was generated by summing the RF data along one direction, and the central slice image was taken from the volumetric image, both followed by the same method of SVD filtering processing. Once the MB-only datasets were acquired, an initial noise threshold was applied to remove the noise, then centroiding together with the adaptive thresholding method was used to localize and isolate MBs. An initial threshold was applied to remove the noise and each frame was segmented into small patches. The number of bubbles in each patch was justified using parameters including size, intensity, solidity and eccentricity of the connected regions. If more than one bubble is considered to be within the patch, threshold will be increased to segment this patch into smaller patches until multiple bubble patches are disappeared. The final localization image was plotted by accumulating these centroids and smoothed using a Gaussian kernel.

\section{RESULTS}
\subsection{Simulation}
\subsubsection{Single Scatter}

\begin{figure*}[h]
\centerline{\includegraphics[width=\textwidth]{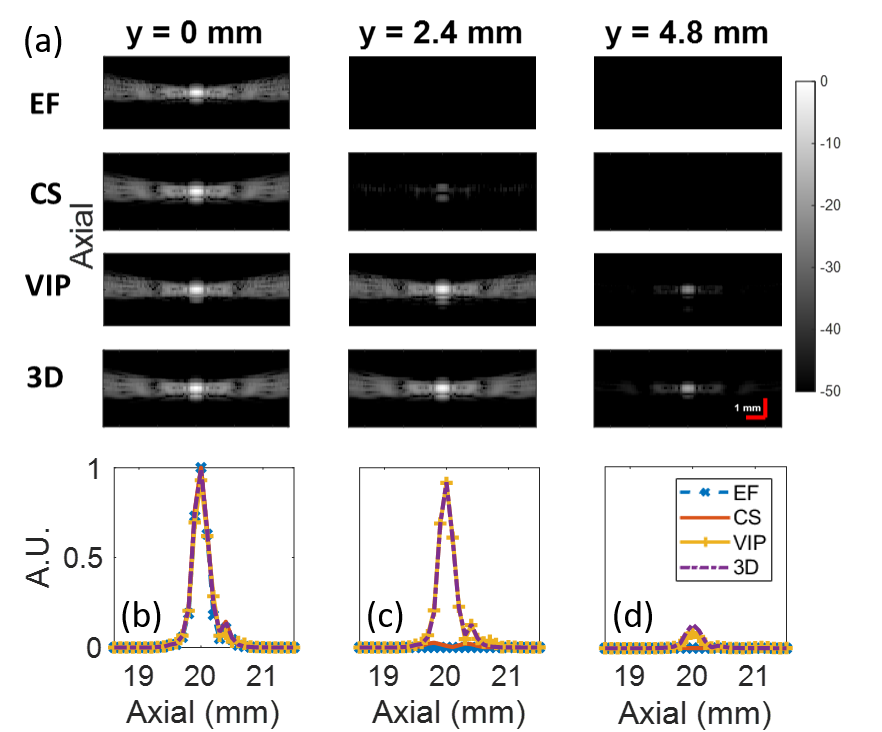}}
\caption{(a) PSFs. The intensity profile along the axial direction when the single scatter is placed at (b) y=0 (left), (c) y= 2.4 (middle) and (d) y=4.8mm (right) respectively. The first row in (a) shows the PSF with an elevational focus (EF) in transmit and receive; The second row in (a) shows the PSF of the central slice (CS) of the volumetric image; the third row in (a) shows the PSF of the volumetric information projection (VIP) method; the fourth row in (a) shows the PSF of the 2D slice corresponding to the position of the scatter. The red bar indicates the length of 1mm. }
\label{fig4}

\end{figure*}

Fig. \ref{fig4} shows the PSFs when the single scatter in each frame is placed at different elevational positions. The images were normalized by the maximum intensity and then log compressed. Both the 3D and VIP methods can detect the scatters across the volume, while the focused methods cannot. Specifically, in Fig. \ref{fig4}(a), the first row shows that an elevational focus in transmit and receive omits scatters that are not within the central imaging plane. In the second row, where the 2D image slices correspond to the central slice of the whole imaging volume, some side lobe signals are visible when y=2.4 mm. In the VIP method, the scatter is visible even if it is placed at the boundary of the imaging volume (y=4.8 mm). The last row shows a 2D slice of the 3D volume, different from the CS scenario, this slice was chosen at exactly where the scatter was, which means that it is the central slice of the main lobe of the 3D PSF rather than the whole imaging volume. It can be seen that VIP is comparable to the 3D method in capturing the scatters within the 3D volume. Fig. \ref{fig4}(b) shows the axial intensity profiles for different datasets. It shows that only VIP and 3D methods are able to detect off-center plane scatters, although the amplitude of the signal varies significantly depending on the scatter locations.

\begin{figure}[h]
\centerline{\includegraphics[width=0.7\columnwidth]{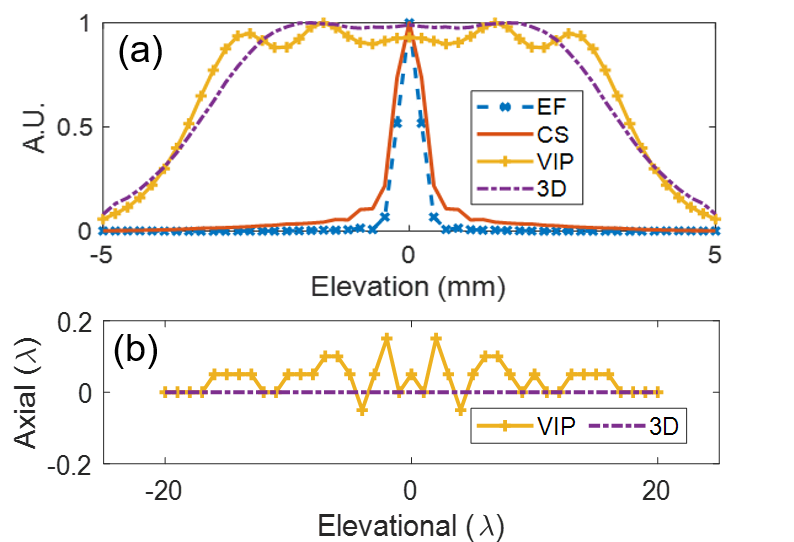}}
\caption{(a) Imaging sensitivity in the elevational direction for the four imaging methods. (b) Localization errors along the axial direction.}
\label{fig5}
\end{figure}

Fig. \ref{fig5}(a) shows the peak intensity when the single scatter is placed at the elevational focus depth (x = 0, z = 20) mm but at different elevational positions. It can be seen that the peak intensity profile decreased sharply when the scatter is slightly out-of-plane in both the EF and CS methods, while the VIP and 3D methods have a similar contour of profile. Fig. \ref{fig5}(b) shows the centroid localization errors in the unit of wavelength of the four imaging methods. The lateral localization errors are all close to zero for these four methods. The axial localization errors for the EF and CS methods are also not shown in this figure as any scatter beyond the central region cannot be detected as their intensities fall below the initial noise threshold.

It is reasonable as the PSF along the lateral direction is almost symmetrical and thus the centroid calculated is less likely to be shifted away from the true position. It is worth noticing that the localization error of the VIP method is larger than the 3D method and is fluctuating around zero as the scatter position varies along the elevational direction. 3D localization performed well in both directions.

\subsubsection{Cross-Tube Phantom}
\begin{figure}[h]
\centerline{\includegraphics[width=0.8\textwidth]{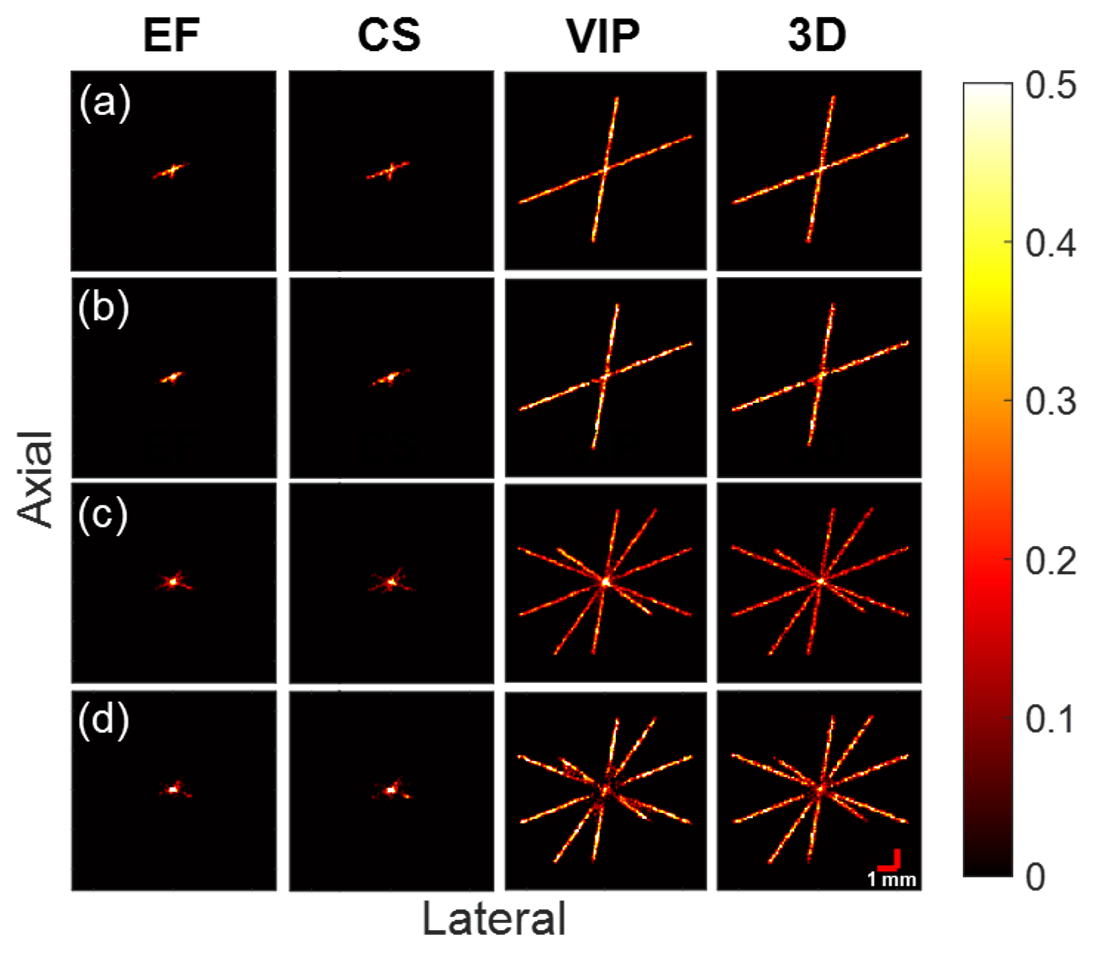}}
\caption{Localization images of the (a) (b) 2-tube and (c) (d) 5-tube phantom, with (a) (c) lower scatter concentration (l.c.) and (b) (d) higher scatter concentration (h.c.), produced by using the four methods: Elevational Focused (EF), Central Slice (CS), VIP, and 3D.}
\label{fig6}
\end{figure}

Fig. \ref{fig6} shows the localization images for the 2-tube and 5-tube phantoms with different scatter concentrations. Fig. \ref{fig6}(a) and (c) show the results of low scatter concentration when each frame has 1 scatter per microtube, i.e. there are 2 and 5 scatters present at each frame in total in the 2-tube and 5-tube phantoms, respectively. With the ground truth shown in Fig. \ref{fig2}. 

\begin{figure}[!b]
\centerline{\includegraphics[width=0.8\textwidth]{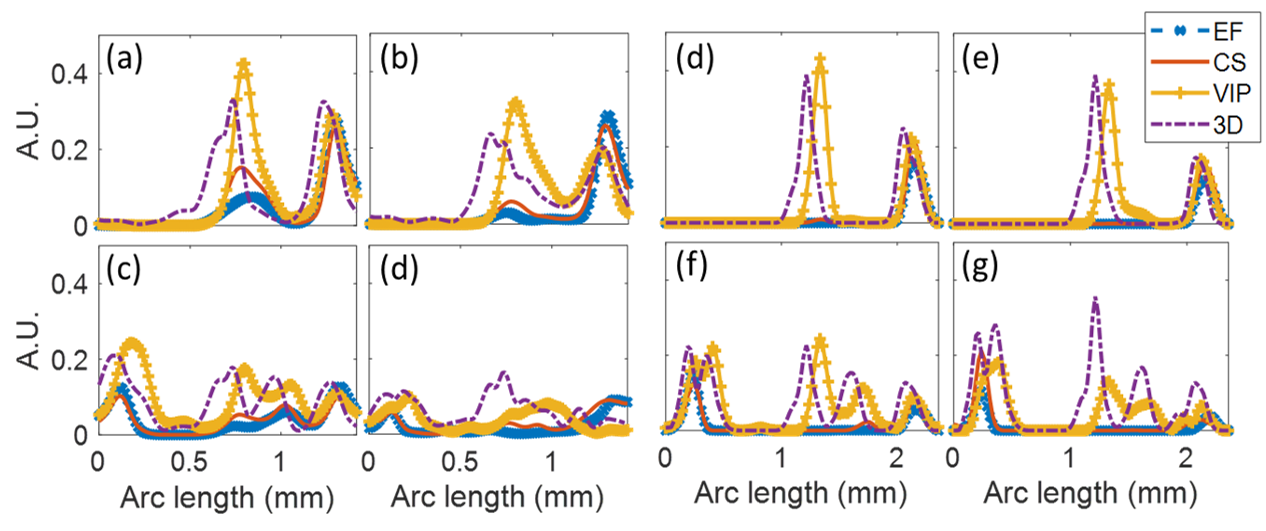}}
\caption{Radial intensity of the SR images along the two black rings as shown in Fig. \ref{fig2}(g). Top row: two tube results; Bottom row: five tube results; (a)(c) inner ring with low scatter concentration; (b)(d) inner ring with higher scatter concentration; (d)-(g) the same as (a)-(d)}
\label{fig7}
\end{figure}

It can be seen that both the EF and CS methods can only image the central part of the phantom, the fourth column shows the maximum intensity projection (MIP) of the 3D localization results on the same imaging plane as the other three imaging methods. It can be seen that the volumetric information captured by VIP-SR is comparable to the 3D SR method. As the scatter concentration increases in Fig. \ref{fig6}(b) and (d), where each frame has 5 scatters per microtube, it is shown that the majority part of the microtube structures is missing in the EF and CS method, while most of them are visible in the VIP and 3D method.

It is worth noticing that in Fig. \ref{fig6}(d), when both the structure complexity and scatter density increased, both the VIP and 3D methods witnessed an image deterioration, shown as a dark central region where the microtubes get closer. When the scatter concentration is very high, all the imaging methods failed in resolving the structures in the center. In Fig. \ref{fig7}(f) and (g), the intensity curve of the VIP-SR method saw a more prominent intensity reduction than the 3D SR curve compared to the low concentration simulation results in Fig. \ref{fig7}(d) and (e), which indicates that the VIP-SR method is more vulnerable to higher bubble concentrations and complex micro-vessel structures.

\begin{table}[H]
\caption{Evaluation Parameters}
\label{table}
\renewcommand{\arraystretch}{1.3}
\centering
\begin{tabular}{p{40pt}p{40pt}p{20pt}p{20pt}p{20pt}p{20pt}}
\toprule
                                                 &             & \multicolumn{4}{c}{Methods} \\ \cmidrule(l){3-6} 
\multirow{-2}{*}{Phantom}                                    & \multirow{-2}{*}{Metric} & \textbf{EF} & \textbf{CS} & \textbf{VIP} & \textbf{3D}  \\ \midrule
                                                 & Precision        & 0.98 & 0.98 & 0.99 & 0.99 \\
                                                 & Sensitivity       & 0.12 & 0.09 & 0.97 & 0.98 \\
\multirow{-3}{*}{\begin{tabular}[c]{@{}c@{}}2-tube\\ l.c.\end{tabular}}             & J.I.           & 0.12 & 0.09 & 0.96 & 0.98 \\
\rowcolor[HTML]{EFEFEF} 
\cellcolor[HTML]{EFEFEF}                                     & Precision        & 0.84 & 0.84 & 0.81 & 0.78 \\
\rowcolor[HTML]{EFEFEF} 
\cellcolor[HTML]{EFEFEF}                                     & Sensitivity       & 0.06 & 0.07 & 0.57 & 0.54 \\
\rowcolor[HTML]{EFEFEF} 
\multirow{-3}{*}{\cellcolor[HTML]{EFEFEF}\begin{tabular}[c]{@{}c@{}}2-tube\\ h.c.\end{tabular}} & J.I.           & 0.06 & 0.07 & 0.50 & 0.47 \\
                                                 & Precision        & 0.92 & 0.87 & 0.94 & 0.98 \\
                                                 & Sensitivity       & 0.10 & 0.05 & 0.86 & 0.93 \\
\multirow{-3}{*}{\begin{tabular}[c]{@{}c@{}}5-tube \\ l.c.\end{tabular}}             & J.I.           & 0.10 & 0.05 & 0.81 & 0.91 \\
\rowcolor[HTML]{EFEFEF} 
\cellcolor[HTML]{EFEFEF}                                     & Precision        & 0.62 & 0.56 & 0.78 & 0.80 \\
\rowcolor[HTML]{EFEFEF} 
\cellcolor[HTML]{EFEFEF}                                     & Sensitivity       & 0.03 & 0.03 & 0.43 & 0.50 \\
\rowcolor[HTML]{EFEFEF} 
\multirow{-3}{*}{\cellcolor[HTML]{EFEFEF}\begin{tabular}[c]{@{}c@{}}5-tube \\ h.c.\end{tabular}} & J.I.           & 0.03 & 0.03 & 0.38 & 0.45 \\ \bottomrule 
\end{tabular}
\label{tab2}
\end{table}

In Table \ref{tab2}, the sensitivity of the VIP method saw a considerate increase compared to the EF and CS method and is close to the 3D method, notwithstanding the scatter concentration and structure complexity, although an abrupt drop in the sensitivity and Jaccard Index (J.I.) as the scatter concentration increased can be noticed.

\subsection{\textit{In vitro} Experiments}

\begin{figure*}[!t]
\centerline{\includegraphics[width=\textwidth]{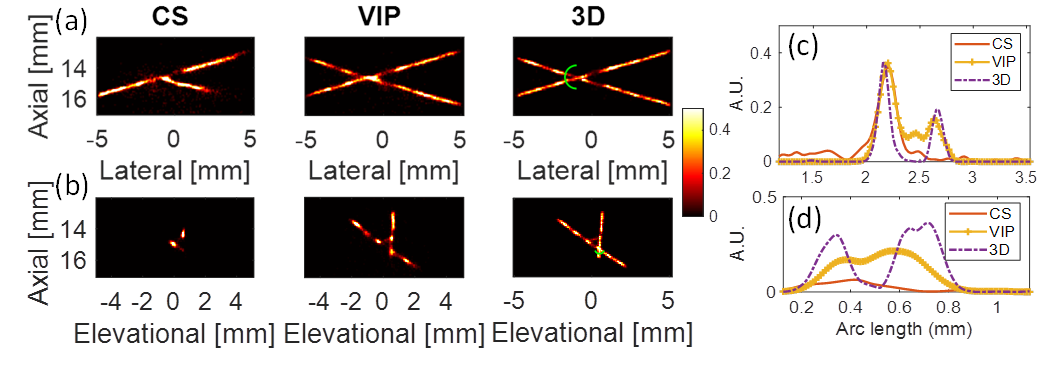}}
\caption{\textit{In vitro} localization images of different imaging methods along (a) elevational and (b) lateral direction. 3D Results are the maximum intensity projection images. Green circles indicate radius of 750 $\mu$m in (a) and 400$\mu$m in (b).}
\label{fig8}
\end{figure*}

Fig. \ref{fig8} shows the localization images of different imaging methods with the microvascular flow phantom shown in Fig. \ref{fig3}. In the CS imaging method, only parts of the 3D tube phantom can be seen, whereas the localization image of the VIP method shows a similar structure as the MIP of the 3D result, at a cost of less resolving ability, which is confirmed by the radial intensity profile along the half green circle in Fig. \ref{fig8}(c) and (d).

\begin{figure}[!b]
\centerline{\includegraphics[width=\columnwidth]{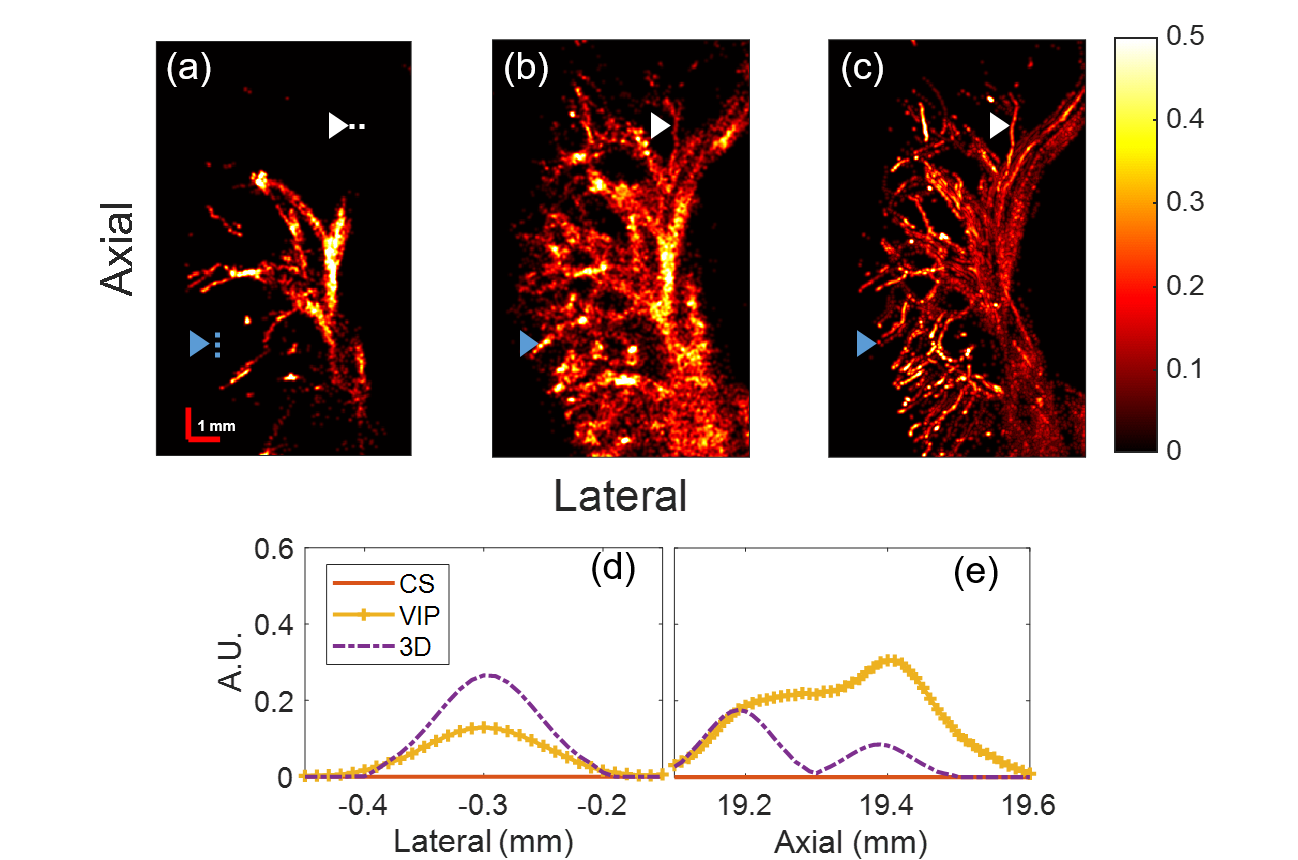}}
\caption{\textit{In vivo} Localization images of the (a) central slice, (b) VIP and (c) MIP of the 3D imaging methods. Intensity profile of the localization image along the (d) white horizontal and (e) blue vertical dashed lines as indicated in (a).}
\label{fig9}
\end{figure}

\subsection{\textit{In vivo} Experiments}

Fig. \ref{fig9} shows the localization images of the left rabbit kidney using different imaging methods. In \ref{fig9}(d), a limited number of microbubbles can be localized and seen, whereas the localization image of the VIP method shows a much denser network of microvessels. The results and the level of detail of the VIP method are very similar to the maximum projection image of the 3D acquisition. The white triangle indicates the particular advantage of the VIP method, When the vessel network stretches out from the central imaging plane, it can no longer be reconstructed with the CS method. The VIP method on the other hand successfully captures the out-of-plane microvasculature. This is further confirmed by the 3D result and the intensity profile in Fig. \ref{fig9}(d), where the intensity profile of the CS method is flat, and both the VIP and 3D methods see peak profiles. 

\section{DISCUSSION}
SRUS/ULM with volumetric information is valuable but existing techniques using matrix array transducers face significant challenges in both the matrix array fabrication complexity and capability in data transfer and processing. Inspired by clinical X-Ray radiography, this study demonstrates that it is feasible to obtain volumetric micro-vessel information through VIP-SR using a simple 1D linear transducer array without an elevational focus. In this proof-of-concept study, we have demonstrated in both simulation and experiments that VIP-SR can visualize 3D vascular structures, the majority of which are not visible using a classic 1D linear array that has an elevational focus. Such an approach could significantly expand the field of view of SRUS in the third dimension without any additional hardware cost and data transfer/processing cost compared to the conventional 2D imaging method. 

Table \ref{tab3} shows the comparison of the different imaging methods when processing one \textit{in vivo} dataset. Compared with the 3D imaging method, the VIP method only needs 32 channels of RF data, a 32 folds decrease in transducer complexity, 10 times less beamforming time, 140 times less memory space, and 360 times less SVD processing time. This simplicity will facilitate the clinical translation of VIP-SR.

\begin{table}[h]
\caption{Computational Cost}
\label{table}
\renewcommand{\arraystretch}{1.3}
\centering
\begin{tabular}{lll}
\toprule
                                                                                       & \textbf{CS/VIP}                                                        & \textbf{3D}                                                                  \\ \midrule
Channel   Number                                                                       & 32                                                                     & 1024                                                                         \\
\begin{tabular}[c]{@{}l@{}}Image Size   \\ (100 $\mu$m isotropic pixel/voxel)\end{tabular} & \begin{tabular}[c]{@{}l@{}}131*81*1500   \\ (z, x, frame)\end{tabular} & \begin{tabular}[c]{@{}l@{}}131*81*143*1500  \\ (z, x, y, frame)\end{tabular} \\
Memory Size                                                                            & 127.3 MB                                                                & 18.2 GB                                                                       \\
\begin{tabular}[c]{@{}l@{}}Beamforming time \\ (on NVIDIA   TITAN Xp)\end{tabular}     & 2 mins                                                                 & 25 mins                                                                      \\
\begin{tabular}[c]{@{}l@{}}SVD processing time \\ (on 128GB RAM)\end{tabular}          & 20 seconds                                                             & 2 hours                                                                      \\ \bottomrule
\end{tabular}
\label{tab3}
\end{table}

Fig. \ref{fig4}(c)  shows that the intensity profile of the VIP and 3D method almost overlapped for all the scatter positions, while the other two methods can only pick up the in-plane scatter. In Fig. \ref{fig4}(d), it can be seen that when the scatter reaches the edge, all the intensities dropped significantly, but the 3D method and VIP can still detect the signals, while the 3D method performs slightly better than VIP in preserving the intensities of side. 

Our study has also investigated the impact of vascular complexity and microbubble concentration on the SR US images. While the image quality of both 3D SR and VIP-SR decreases as the vascular complexity and bubble concentration increase, the VIP-SR is more susceptible to these factors (Fig. \ref{fig6}(d)). This can also be confirmed by the \textit{in vivo} results. In Fig. \ref{fig9}, the CS imaging method captured much fewer in-plane vessels than 3D and VIP-SR. While 3D and VIP-SR seem to show similar volumetric coverages (Fig. \ref{fig6}(b)(c)), the VIP method shows lower resolving ability, which is likely due to the overlapping signals in the elevational direction obscure each other and impacting the localization. Therefore VIP-SR would be ideal for imaging tissues with sparse vasculature, and for tissue with denser vasculature, a compromise between bubble concentration and acquisition time is needed to ensure the bubble signals are well separated to ensure the SR image quality. Furthermore, the systematic localization errors shown in Fig. \ref{fig5}(b) may also cause some differences in the vascular structures between the VIP-SR and the projection image of full 3D SRUS results.

It should be noted that even with the existing 1D array SRUS imaging, there is a certain amount of volumetric information in the images already, as there is always a slice thickness of the 2D images. However, by using an unfocused lens in VIP-SR such a field of view can be significantly expanded as demonstrated in this study. It is also possible to use a diverging lens to further expand the field of view \cite{Favre2022}.

The elevational size of the 1D transducer aperture determines the size of the VIP imaging volume being covered, but increasing the aperture size can compromise the localization performance in the near field. As the elevational aperture increases, the time-of-arrival of the signals across each long element would be more variable in the near field and cause both constructive and destructive interference, resulting in the elongation of the PSF along the axial direction and reduction of its amplitude. In Fig. \ref{fig5}we have already demonstrated that localization errors exist with VIP-SR and the errors oscillate as a function of the elevational position, which is likely due to the coherence interference of bubble signals arriving at the different parts of the same long transducer element over time. As the scatter moves away from the center imaging plane, the first-time-of-arrival of the ultrasound wave on different elevational positions would have larger deviations, and as a result of superimposition along this direction, the summed RF data would have longer ripples along the axial direction, shifting the weighted centroid backward as a result. Such elongated weaker PSF may also be difficult to recognize by bubble detection algorithms. However, our results show that the maximum error along the axial direction using centroiding is no more than 0.2 lambda, which is still well below the diffraction limit. Furthermore, it is worth mentioning that, this axial localization error will decrease as the imaging depth increases when the elevational shift is not comparable to it, while errors will increase in the near imaging field as the transmit and receive path will be more dominated by the elevational shift. And this may be further addressed by alternative localization methods using cross-correlation \cite{Song2017}, sparsity-based method \cite{Bar-Zion2018} or deconvolution \cite{Yu2018,Yan2022} if the system PSF can be fully characterized.

Microbubble dynamics simulation \cite{Versluis2020} would be desirable but not essential in this proof-of-concept study as the aim of the study is to investigate the spatial coverage of the proposed VIP-SR imaging method. Furthermore, the performance of the VIP method in velocity tracking was not investigated. While in principle, the in-plane motion can still be tracked, it is difficult to estimate any out-of-plane flow in the projected image. And overlapping of the projected structures would also cause ambiguity in the tracking process. However, we have shown that the PSF pattern varies at different elevational positions, and it may be possible to use this feature to infer flow velocity in the elevational direction. Future studies are required to investigate the feasibility.

It should be noted that in the experimental results, the brightness of the microtube phantom was varying significantly along the tube, which is likely caused by the non-uniform pressure field in this direction due to the existence of gaps between the four sub-apertures of the Vermon matrix transducer. 

\section{CONCLUSION}
In summary, this study demonstrates that VIP-SR can generate super-localization microscopy images that capture volumetric information of micro-vasculature using a linear 1D array probe without an elevational focus lens. The technique requires hardware and computing resources that are many folds less than the full 3D SRUS and similar to that required by a classical 2D imaging technique.

\section*{Acknowledgment}
We would like to thank Dr. Jacob Broughton-Venner for the discussion and Mr. Qingyuan Tan for the help with the plot of figures.

\bibliography{VIPSR}
\end{document}